\documentclass[doublecol,linenumbers]{epl2}
\usepackage{graphicx}
\def\dbar{{\mathchar'26\mkern-12mu d}}
\usepackage{epstopdf}
\usepackage{amsmath}
\usepackage{amssymb}
\usepackage{amsfonts}

\title{Single-ion quantum Otto engine with always-on bath interaction}
\author{Suman Chand \and Asoka Biswas}
\institute{%
Department of Physics, Indian Institute of Technology Ropar, Rupnagar, Punjab - 140001, India
}%
\date{\today}%

\abstract{
We demonstrate how the reciprocating heat cycle of a quantum Otto engine (QOE) can be implemented using a single ion and an always-on thermal environment. The internal degree of freedom of the ion is chosen as the working fluid, while the motional degree of freedom can be used as the cold bath. We show, that by adiabatically changing the local magnetic field, the work efficiency can be asymptotically made unity. We propose a projective measurement of the internal state of the ion that mimics the release of heat into the cold bath during the engine cycle. In our proposal, the coupling to the hot and the cold baths need not be switched off and on in an alternate fashion during the engine cycle, unlike other existing proposals of QOE. This renders the proposal experimentally feasible using the available tapped-ion engineering technology.}


\begin{document}
\maketitle

\section{\label{sec:i}Introduction}
Quantum heat engines (QHE) have recently attracted great interest since it was originally proposed in \cite{scovil}. These engines employ a quantum system as the working substance and like their classical counterparts, are expected to run cyclically between two heat baths. It absorbs $\left(Q_{H}\right)$ heat from the hot bath, performs certain mechanical
work $W$, and rejects heat $Q_{L}$ to the cold bath.  While a classical Carnot engine cannot extract work when the temperatures of the two baths become equal (pertaining to a single bath), it is shown to be possible using the quantum property of the working substance. Scully and his coworkers have shown that it is possible to extract work from a single bath using quantum coherence \cite{Scully-science,scully-PRL2} or quantum negentropy \cite{Scully-PRL-negentropy}, with certain unique features of the QHE, when compared to their classical counterparts. It is not even necessary to maintain the thermal equilibrium of the quantum baths \cite{quan} to extract work using a QHE.

Several proposals for implementing quantum heat engines have been made using spin systems \cite{kosloff,he,wu,quan,wang1}, harmonic oscillators \cite{quan,wu,rezek}, and multi-level systems \cite{bender,henrich,abe,quan1}. In \cite{kosloff}, an ensemble of non-interacting two-level systems, driven by a local external field and subject to frictional force, is coupled to the baths for {\it finite} time and the optimized time-scale of engine operation has been derived. In \cite{must1,johal}, the interacting spins have been considered and the adiabatic processes have been performed by changing the local magnetic fields that drive the spins.  Correspondence of the efficiency of such coupled-spin system to the entanglement has been studied for Heisenberg interaction \cite{zhang,albayrak,must2} and Dzyaloshinski-Moriya interaction \cite{zhang1}. Comparisons between quantum versions of Carnot engine and Otto engine have been studied in \cite{wang} for a two-level system as well as for harmonic oscillators. It is further shown \cite{hardal}  that photons in an optical cavity can also be used as a working substance for a QHE in which a coherent cluster of atoms serves the role of `fuel' through their superradiance. However, in all these proposals, the system undergoes a heat cycle, during which it interacts with a hot bath and a cold bath (both modelled as a classical system) in an alternative fashion. It is assumed that one can switch off the interaction with the bath during a certain stage of the heat-cycle, the so-called reciprocating cycle. For a quantum system, however, such an interaction is always on \cite{kurizki,kurizki1} and it is experimentally challenging to turn it off or on during the heat cycle. Dynamical decoupling from the bath by applying a certain pulse sequences \cite{bang} to the system may serve the purpose. But, such a technological requirement can be overwhelming so as to mimic the classical heat cycle in a quantum system.


In this paper, we propose an experimentally feasible model of a single-ion QHE, that works like a reciprocating heat cycle, but maintaining the always-on interaction with the bath, like in the continuous heat cycles \cite{kurizki,kurizki1}. The electronic degree of freedom of the ion is chosen as the working substance S, while its vibrational degree of freedom plays the role of the cold bath, that also interacts with the ion throughout the entire cycle. The thermal environment here acts as the hot bath. We show that while the hot bath thermalizes the state of S, the heat transfer to the `cold bath' can be achieved by a projective measurement of the states of S. We emphasize that the fact that the interaction with the baths is never switched off during the entire cycle, makes our model feasible in experiments. 
We show that the efficiency of this engine can be made close to unity by manipulating a local magnetic field (that works as a ``piston") applied to the ion. Note that the proposal of implementing QHE using a single ion exists \cite{abah}, in which the frequency of the linear Paul trap is changed during the isentropic processes of the heat engine, while during isochoric processes, the system is weakly coupled to the hot or cold bath to achieve the thermal equilibrium at the bath temperature. This model clearly requires alternative coupling to the baths. This proposal has been implemented using $^{40}$Ca$^+$ ion with an efficiency 1.9\% at the maximum power limit \cite{ion-expt}. On the contrary, in our model, one only requires to change the driving magnetic field adiabatically and a projective measurement of the electronic states, both of which can be routinely achieved in a trapped-ion set up \cite{ion-review}.

The paper is organized as follows. In Sec. II, we discuss the QHE model based on a single trapped ion. We present all the relevant Hamiltonians and the achievable efficiency in this QHE. In Sec. III, we describe all the required stages of the heat cycles. We also present the conditions that are relevant to successfully implement these stages. We conclude the paper in Sec. IV.

\section{Model}\label{s:ii}
In this paper, we focus on implementing a quantum Otto engine (QOE). A QOE cycle consists of four stages: a) In ignition stroke (an isochoric process) the system S gets thermalized by absorbing $Q_H$ heat from a hot bath at a temperature $T_H$. b) In the next stroke (the `expansion' stroke), the system undergoes an adiabatic process, thereby maintaining the thermal equilibrium, such that any kind of exchange of heat with the bath is inhibited. During this process, an external work $W$ is done by the system amounting to an adiabatic change of a local driving field (the `piston'). c) In the following stroke, the `exhaust' stroke (an isochoric process), the system releases $Q_L$ energy to a cold bath. d) In the last stroke (the `compression' stroke), the system initializes itself through an adiabatic evolution, that also initializes the local driving field. In the following, we will describe how to implement all these strokes using a trapped ion.

A single trapped ion in Lamb-Dicke limit can be considered as a two-qubit coupled system [see Fig. 1], in which the ion is confined in its two lowest lying internal states $|g\rangle$ and $|e\rangle$ (represented by the relevant Pauli matrices $\sigma_{x,y,z}$) and the lowest lying vibrational states $|0\rangle$ and $|1\rangle$ (such that $a^\dag |1\rangle$ vanishes, where $a^\dag$ is the creation operator of the vibrational mode). The Hamiltonian that describes the interaction between the internal and motional states of the ion  can thus be written as (in unit of Planck's constant $\hbar=1$)

\begin{equation}
H_1=H_S+H_{\rm ph}+H_{\rm int}\;,
\end{equation}
where

\begin{eqnarray}
&&H_S=g\sigma_{x}+B\sigma_{z}\;,\;H_{\rm ph}=\omega a^{\dagger}a\;,\\
&&H_{\rm int}=k\left(a^{\dagger}\sigma_{-}+\sigma_{+}a\right)\;.\label{eq:1}
\end{eqnarray}
Here, $H_S$ is the Hamiltonian for the internal states of the ion, $H_{\rm ph}$ represents the energy of the phonons, relevant to the vibrational degree of freedom, and  $H_{\rm int}$ defines the interaction between the internal and the vibrational degrees of freedom of the ion. We consider that the internal states are driven by a local electric field with Rabi frequency $2g$ and a magnetic field of strength $B$, applied along the quantization axis. The vibrational frequency of the ion is chosen as $\omega$ and $k$ represents the interaction between the internal states and the motional states of the ion.

In context of the QOE, we consider the internal degree of freedom as the working substance S. The eigenvalues of the relevant Hamiltonian $H_S$ are given by $E_{1,2}=\pm \sqrt{g^2+B^2}$, with the respective eigenstates

\begin{eqnarray}
\label{en}|E_{1,2}\rangle &=&\frac{1}{N_\mp}\left[\left(B\mp \sqrt{g^2+B^2}\right)|g\rangle - g|e\rangle\right]\;,\\
\label{npm}N\mp &=& \sqrt{2\left[g^2+B^2\mp B\sqrt{g^2+B^2}\right]}\;.
\end{eqnarray}

\subsection{Efficiency of a two-level QOE}
Following Kieu \cite{kieu}, we can now calculate the efficiency of a QOE based on the above two-level system S. The average energy of this system can be written as $U=\sum_n E_n P_n$ ($n=1,2$), where $P_n$ is the occupation probability of the energy eigenstate $|E_n\rangle$. Comparing $dU=\sum_n E_n dP_n + \sum_n P_n dE_n$ with the first law of classical thermodynamics, $dU=\dbar Q+\dbar W$ ($\dbar$ defines non-exact differential and refers to the path-dependence, see e.g., \cite{zeem,goold}), one can identify in the quantum regime the infinitesimal
heat transfer as $\dbar Q=\sum_n E_n dP_n$ and the infinitesimal work done as $\dbar W=\sum_n P_n dE_n$. It is to be noted that this expression of $\dbar Q$ is valid irrespective of whether the system is in thermal equilibrium or not \cite{quan}, contrary to its classical counterpart $\dbar Q=TdS$. This also includes the contribution of the ergotropy and adiabatic work \cite{goold}. Clearly, the heat transfer is associated with the change in the probability distribution for various eigenstates, while the work done is related to change in the energy eigenvalues, keeping the probability distribution the same. For the two-level system S, the heat absorbed $Q_H$ during the ignition stroke can thus be written as

\begin{equation}\label{qh}
Q_H=\sum_{n=1}^2E_n^H\left[P_n^H-P_n^L\right]\;,
\end{equation}
where $E_n^H$ is the $n$th energy eigenvalue while the system interacts with the hot bath at the temperature $T_H$, $P_n^L$ is the initial probability of the $n$th energy eigenstates and $P_n^H$ is that after thermalization. In a similar way, the heat released $Q_L$ during the exhaust stroke can be written as

\begin{equation}\label{ql}
Q_L=\sum_{n=1}^{L}E_n^L\left[P_n^L-P_n^H\right]\;,
\end{equation}
where $E_n^L$ is the $n$th energy eigenvalue while the system interacts with the cold bath. Note that during the adiabatic processes, the probabilities $P_n^H$ and $P_n^L$ do not change. However the energy eigenvalues $E_n$ change, due to the adiabatic change in the local driving field. The work done during the heat cycle can thus be written as

\begin{equation}\label{work}
W=Q_H-|Q_L|=\sum_{n=1}^{2}\left(E_n^H-E_n^L\right)\left[P_n^H-P_n^L\right]\;,
\end{equation}
leading to the following expression of the work efficiency $\eta=\frac{W}{Q_H}$:

\begin{eqnarray}\label{eff}
\eta &=&\frac{\sum_{n=1}^{2}\left(E_n^H-E_n^L\right)\left[P_n^H-P_n^L\right]}{\sum_{n=1}^2E_n^H\left[P_n^H-P_n^L\right]}\;,\nonumber\\
&=&\frac{(E_1^H-E_2^H)-(E_1^L-E_2^L)}{E_1^H-E_2^H}\nonumber\\
&=&1-\frac{E_1^L-E_2^L}{E_1^H-E_2^H}
\end{eqnarray}
where we have used the fact $\sum_{n=1}^2P_n^H=\sum_{n=1}^2P_n^L=1$.
If the local driving fields are changed from $B=B_H$ and $g=g_H$ to $B=B_L$ and $g=g_L$, respectively, during the expansion stroke, the efficiency takes the following form:

\begin{equation}\label{eff-fin}
\eta=1-\sqrt{\frac{g_L^2+B_L^2}{g_H^2+B_H^2}}\;.
\end{equation}
This suggests that in the limit of $B_L,g_L\rightarrow 0$, one could ideally achieve the efficiency close to unity. In practice, the local electric field that drives the ion can be kept constant at a small value of $g$, such that the larger efficiency can be obtained by manipulating only the magnetic field.   
The adiabatic process is performed by adiabatic changes of the magnetic field from $B_H$ to $B_L (<B_H)$ during the expansion stroke and from $B_L$ to $B_H$ during the compression stroke. Such a control of the efficiency by a local magnetic field has also been proposed in \cite{johal}. The ability to control the electric and the magnetic fields independently makes the model more flexible towards achieving larger efficiency.

\section{Implementation of the QOE cycles}
As outlined before, we consider the internal degree of freedom as the working substance S. On the other hand, in the Lamb-Dicke limit, the ionic motion is confined to its two lowest lying states, while the higher excited states are not populated. For example, a single Be ion can be cooled using standard ion trapping technique, such that the average motional quantum number can be of the order of 0.02 (see for example, \cite{monroe}). In this way, the motional states can be considered as the relevant two-level cold bath, such that average phonon number in the vibrational degree of freedom $\bar{n}_{\rm ph}\ll 1$.  Note that the system S always interacts with this effective cold bath through the Hamiltonian $H_{\rm int}$. We here emphasize that a finite-level system can act as a bath, as coupling to such a bath often leads to decoherence of the system (see, e.g., \cite{asoka}). The thermal environment at an equilibrium temperature $T_H$ also interacts with the system S and the ionic motion.

\begin{figure}
\includegraphics[width=8cm,height=6cm]{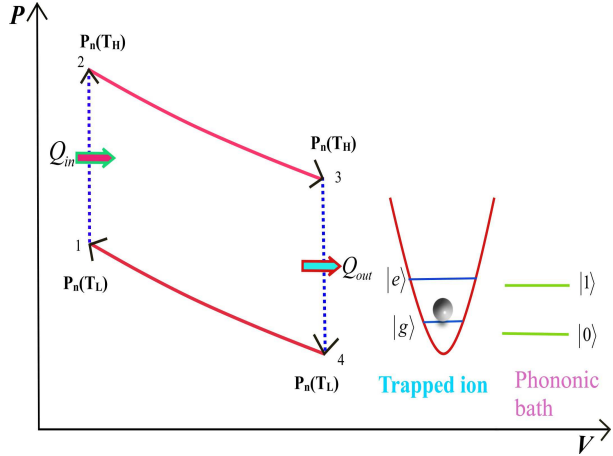}
\caption{Schematic diagram of the trapped ion QOE. The red (blue) lines refer to the adiabatic (isochoric) processes. The insets display the relevant energy levels of the internal and the vibrational degrees of freedom, relevant to the ionic motion.}
\end{figure}

The QOE consists of four different strokes - two isochoric strokes and two adiabatic strokes. In the following, we describe how to implement all these strokes with the system S and the two baths identified as above.

\subsection{Ignition Stroke}
During this isochoric process ($1\rightarrow 2$, see Fig. 1), the ion interacts with the hot bath and gets thermalized through Markovian evolution. This leads to the following mixed states of the system S and the phonons:

\begin{equation}\label{ig1}
\rho_1^{(H)}=\sum_{i=1}^4p_i|U_i\rangle\langle U_i|\;,\;p_i=\frac{\exp\left(-U_i/k_BT_H\right)}{\sum_{i=1}^4\exp\left(-U_i/k_BT_H\right)}\;,
\end{equation}
where $p_i$ is the occupation probability of the $i$th eigenstate $|U_i\rangle$ of the total Hamiltonian $H_1$ for the (S+phonon) system.
To identify the eigenvalue $U_i$, we rewrite the Hamiltonian $H_1$ in the joint basis $\left\{ \left|g,0\right\rangle ,\left|g,1\right\rangle ,\left|e,0\right\rangle ,\left|e,1\right\rangle \right\}$ of the internal states and the motional states in the following matrix form:

\begin{equation}
H_{s,ph}=\left(\begin{array}{cccc}
-B & 0 & g & 0\\
0 & -B+\omega & k & g\\
g & k & B & 0\\
0 & g & 0 & B+\omega
\end{array}\right).
\end{equation}
The eigenvalues of this Hamiltonian are given by

\begin{equation}
U_{1,2}  = \frac{1}{2}\left(\omega\mp A_-\right)\;,\;\;U_{3,4} =\frac{1}{2}\left(\omega\mp A_+\right)\;,
\end{equation}
where

\begin{eqnarray*}
A_\pm&=&\sqrt{C\pm 2D}\;,\;\; C = 4B^{2}+4g^{2}+2k^{2}+\omega^{2},\\
 D &=& \sqrt{4g^{2}k^{2}+k^{4}-4Bk^{2}\omega+4B^{2}\omega^{2}+4g^{2}\omega^{2}}\;.
\end{eqnarray*}

Here $|U_i\rangle$ can be written in terms of the joint basis of the internal and motional states as

\begin{equation}
|U_i\rangle=a_{1i}|g,0\rangle+a_{2i}|g,1\rangle+a_{3i}|e,0\rangle+a_{4i}|e,1\rangle\;,\;i=1,2,3,4\;.
\end{equation}
Therefore, the reduced density matrix of the system S can be obtained by taking partial trace over the phonon states as

\begin{eqnarray}
\rho_S^{(H)}&=&\frac{1}{P_H}\sum_{i=1}^{4}e^{-U_i/k_BT_H}\left[(a_{1i}^2+a_{2i}^2)|g\rangle\langle g|\right.\nonumber\\
\label{rhos}&+&\left.(a_{3i}^2+a_{4i}^2)|e\rangle\langle e|+(a_{1i}a_{3i}+a_{2i}a_{4i})(|e\rangle\langle g|+{\rm h.c.})\right.\;\nonumber\\
\end{eqnarray}
where $P_H=\sum_{i=1}^4 \exp[-U_i/k_BT_H]$ is a normalization constant.
This can be written in terms of the energy eigenstates $|E_n\rangle$ of the system Hamiltonian through the inverse transformation of the Eq. (\ref{en}):

\begin{equation}\label{geeigen}
|g\rangle=z_{g1}|E_1\rangle-z_{g2}|E_2\rangle\;,\;|e\rangle=z_{e1}|E_1\rangle-z_{e2}|E_2\rangle\;,
\end{equation}
where

\begin{eqnarray}\label{geecoeff}
z_{g1,g2} & = & -\frac{N\mp}{2\sqrt{g^{2}+B^{2}}}\;,\nonumber\\
z_{e1,e2} & = & z_{g1,g2}\frac{B\pm\sqrt{g^{2}+B^{2}}}{g}\;.
\end{eqnarray}

Using (\ref{rhos}) and (\ref{geeigen}), we can have the occupation probability $P_n^H=\langle E_n|\rho_S^{(H)}|E_n\rangle$ of the eigenstate $|E_n\rangle$  as

\begin{eqnarray}
P_1^H& = & \frac{1}{P_H}\sum_{i=1}^4[e^{-U_i/k_BT_H}\{(a_{1i}^2+a_{2i}^2)z_{g1}^2\nonumber \\
&&+(a_{3i}^2+a_{4i}^2)z_{e1}^2+2(a_{1i}a_{3i}+a_{2i}a_{4i})z_{g1}z_{e1}\}\nonumber\\
\label{p1h}
\end{eqnarray}
and

\begin{eqnarray}
P_2^H& = & \frac{1}{P_H}\sum_{i=1}^4[e^{-U_i/k_BT_H}\{(a_{1i}^2+a_{2i}^2)z_{g2}^2\nonumber \\
&&+(a_{3i}^2+a_{4i}^2)z_{e2}^2+2(a_{1i}a_{3i}+a_{2i}a_{4i})z_{g2}z_{e2}\}\;.\nonumber\\
\label{p2h}
\end{eqnarray}

Assuming that the system S is initially prepared in the state $|g\rangle$, the heat absorbed $Q_H$ during this stroke can be calculated using Eq. (\ref{qh}) as

\begin{equation}\label{qh1}
Q_H=\sum_{n=1}^2E_n^H\{P_n^H-z_{gn}^2\}\;.
\end{equation}
It must be reminded that during this stroke, the magnetic field is kept constant at $B=B_H$, leading to the eigenvalues $E_n^H$ of the system Hamiltonian $H_S$ to remain constant.

\subsection{Expansion stroke}
During this stroke ($2\rightarrow 3$, see Fig. 1), the magnetic field is adiabatically changed from $B_H$ to $B_L$. This means that the occupation probabilities of the two eigenstates $|E_{1,2}\rangle$ do not change; however, the corresponding eigenvalues $E_{1,2}^H$ change to the values $E_{1,2}^L=\pm \sqrt{g^2+B_L^2}$. As shown in Sec \ref{s:iii}, the system does not exchange any heat with the heat bath as well as the phonon modes, i.e., $\dbar Q=0$. This leads to reduction of internal energy of the system, when the following work is performed by the system during this stroke (refer to the first law of thermodynamics: $dU=\dbar Q+\dbar W$):

\begin{equation}\label{w1}
W_1=\sum_{n=1}^2P_n^H(E_n^L-E_n^H)\;,
\end{equation}
where $E_{1,2}^H=\pm \sqrt{g^2+B_H^2}$ are the eigenvalues of $H_S$ before the stroke and $P_n^H$ are given by (\ref{p1h}) and (\ref{p2h}). Note that we change only the magnetic field, that does not change the internal state and only leads to a Zeeman shift.

\subsection{Exhaust stroke}
This is an isochoric process, during which the system releases $Q_L$ heat to the cold bath ($3\rightarrow 4$, see Fig. 1) and the system Hamiltonian changes from $H_S(B_H)$ to $H_S(B_L)$. The initial state of the coupled system (S+phonons) at this stage can be written as

\begin{equation}\label{ex1}
\rho_2^{(H)}=U_I\rho_1^{(H)}U_I^\dag\;,
\end{equation}
where $U_I$ is the unitary operator associated with the adiabatic process defined as

\begin{eqnarray}
U_I&=&{\cal T}\exp[-i\int_0^\tau dt' H(t')]\;,\nonumber\\
H(t)&=&H_S(t)+H_{\rm ph}+H_{\rm int}\;,\nonumber\\
H_S(0)&=&g\sigma_x+B_H\sigma_z\;,\;\;H_S(\tau)=g\sigma_x+B_L\sigma_z\;.
\end{eqnarray}
Here ${\cal T}$ represents time-ordering.
The state $\rho_2^{(H)}$ can be written in the joint basis of the internal and the motional states as

\begin{equation}
\rho_2^{(H)}=\sum_{i,j}\rho_2^{(i,j)}|i\rangle\langle j|\;, |i\rangle,|j\rangle\in |g,0\rangle,|g,1\rangle,|e,0\rangle,|e,1\rangle\;.
\end{equation}
Clearly, at thermal equilibrium, the system S is entangled with the phonon state.  The release of heat from the system to the cold bath is equivalent to cooling down the system. This means that the occupation probabilities of the higher excited states of the system would reduce by such heat release. To facilitate the release of heat from the system to the phonon bath, here we propose a projective measurement of the state of the system S. By measuring the state $|g\rangle$ of the system, the density matrix $\rho_2^{(H)}$ gets factorized and can be written as

\begin{equation}
\rho_2^{(H)}|_{\rm meas}=|g\rangle\langle g|\sum_{k,l=0}^1r_{k,l}|k\rangle\langle l|\;,
\end{equation}
where $|k\rangle, k\in 0,1$ are the states of the phonon modes and $r_{k,l}$ are the relevant density matrix elements between the states $|k\rangle$ and $|l\rangle$.
In this way, the system gets decoupled from the cold bath and the measurement essentially purifies the state of the system S. Such purification process is equivalent to cooling down the system through release of heat to the phonon modes. This is also analogous to algorithmic cooling \cite{alg} in NMR quantum computing, in which a pesudopure state can be prepared from a mixed state of an ensemble of nuclear spins embedded in a single molecule \cite{oscar}. We emphasize that the above measurement does not nullify the interaction Hamiltonian between the system and the baths; rather, it only decouples the system from the baths through factorization of the density matrix. 

Through this cooling process, the probability distribution of the eigenstates of the system Hamiltonian $H_S$ change, keeping the corresponding eigenvalues the same, as the local driving fields are kept constant during this stroke.  The heat released from the system to the cold bath can thus be calculated using Eqs. (\ref{qh}) and (\ref{geecoeff}) as

\begin{equation}\label{ql1}
Q_L=\sum_{n=1}^2E_n^L[z_{gn}^2-P_n^H]\;.
\end{equation}

The measurement process, as described above, is evidently probabilistic and depends upon the outcome of the measurement. A suitable alternative way of decoupling of the system from the bath could be to use the non-selective measurement, as described in \cite{meas}. This is based on a sequence of non-selective quantum non-demolition measurement of the state of the system S, at an interval $\gtrsim 1/B_L$.  This leads to decoupling of S from the phonon bath, i.e., $\rho_{\rm S,ph} \rightarrow \rho_S\otimes \rho_{\rm ph}$, if the measurement outcomes, or alternatively, the states of the measuring device are not read or averaged out. Such a process leads to cooling of the system in the Markovian limit and amounts to heating of the cold bath, as discussed above.

\subsection{Compression Stroke}
During this stroke ($4\rightarrow 1$, see Fig. 1), the system again undergoes through an adiabatic process, during which the magnetic field strength is now adiabatically changed from $B_L$ to $B_H$. The system remains in contact with the the hot bath and the vibrational mode - however, as during the expansion stroke, the occupation probabilities of the energy eigenstates $|E_n\rangle$ remain unaltered at the values $z_{gn}^2$. The eigenvalues change from $E_n^L$ to $E_n^H$ due to the change in the magnetic field. This leads to the following work done during this stroke:

\begin{equation}\label{w2}
W_2=\sum_{n=1}^2z_{gn}^2(E_n^H-E_n^L)\;.
\end{equation}
We emphasize that after the compression stroke, the system remains in the ground state. This can be considered as an initialization for the next cycle. Further, the heat transferred to the vibration mode does not eventually accumulate after a few cycles, as in any case,  the system and the vibrational mode get thermalized by the hot bath during the ignition stroke of each cycle and one can keep on reusing the vibrational mode as a cold bath. 

\subsection{Work Efficiency}
We find that the QOE absorbs heat $Q_H$ [Eq. (\ref{qh1})] during the ignition stroke and releases heat $Q_L$ [Eq. (\ref{ql1})] during the exhaust stroke, while it does certain work $Q_H-|Q_L|$ during the two adiabatic strokes. Therefore, the work efficiency of the QOE can be calculated as

\begin{eqnarray}
\eta=\frac{Q_H-|Q_L|}{Q_H}=\frac{\sum_{n=1}^2[E_n^H-E_n^L]\{P_n^H-z_{gn}^2\}}{\sum_{n=1}^{2}E_n^H\{P_n^H-z_{gn}^2\}}\;.
\end{eqnarray}
Moreover, the projective measurement of a qubit has an energy cost $M\le k_BT_H\ln 2$ \cite{abdel,anders}, where the equality sign holds for a maximally mixed state (referring to maximal change in entropy). This further reduces the effective efficiency of the engine to 

\begin{equation}
\eta_M=\frac{Q_H-|Q_L|}{Q_H+M}\;.
\end{equation}
We show in a parametric plot in Fig. 2 how the efficiency $\eta$ and $\eta_M$ (for a maximally mixed state) vary with the work output $Q_{H}-|Q_L|$ in this system. Such a variation suggests that the efficiency is not limited by the heat absorbed $Q_H$ by the system and by increasing the temperature of the hot bath (and thereby increasing $Q_H$), one can obtain a larger efficiency.
\begin{figure}
\includegraphics[width=8cm,height=6cm]{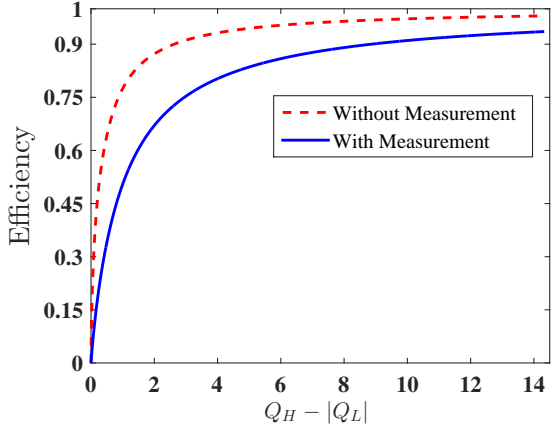}
\caption{Variation of the efficiency $\eta$ (red dashed line) and $\eta_M$ (blue solid line) (for $M=k_BT_H\ln 2$) with the work $Q_H-|Q_L|$ done  by the system, in which the $B_H$ is changed from 0.01 to 10. The other parameters chosen are $B_L=0.01$, $g=0.2$, $k=0.1$, $k_BT_H=1$, and $\omega=1$.}
\end{figure}

\subsection{Adiabaticity}\label{s:iii}
The compression and expansion strokes in an Otto engine are performed adiabatically, during which the system does not share heat with the bath. In the present case, for a quantum version of the Otto engine, one needs to consider quantum adiabatic process. In such a process, the occupation probabilities $P_n$ of the different eigenstates of the system Hamiltonian remain unaltered, though the relevant eigenvalues are adiabatically changed. This is essentially described by $\dbar Q=\sum_n E_ndP_n=0$. To verify that the probabilities $P_n$s do not change during these two strokes, we have studied the dynamics of the system+phonon joint system, using the master equation $\dot{\rho}_{\rm s,ph}=-i[H_{\rm S}(t)+H_{\rm ph}+H_{\rm int},\rho_{\rm s,ph}]$. Here we have assumed that the adiabatic evolution takes place before the time-scale $1/\omega_c$ in which the heat bath, characterized by the temperature $T_H$ becomes effective ($\omega_c$ is the characteristic cut-off frequency of the the bath). We consider a linear variation of the magnetic field, as given by \cite{must2}

\begin{equation}\label{linear}
B(t)=B_H+\frac{B_L-B_H}{\tau}t\;,
\end{equation}
where $\tau$ is the finite time-scale of the change of the magnetic field from $B_H$ to $B_L$ or vice versa. For the initial condition (\ref{ig1}) that the system and the phonon modes are in thermal equilibrium with the heat bath, we solved the above master equation. We show in Fig. 3 that the occupation probabilities of the eigenstates of the system Hamiltonian $H_S$ [see Eqs. (\ref{p1h}) and (\ref{p2h})] do not change substantially for $\tau=5$ (e.g., only a mere $0.1\%$ decrease in the probability $P_1$). This indicates that the system does not exchange heat with both the thermal bath and the phonon modes and thereby the evolution during the $\tau$ interval is adiabatic.

\begin{figure}[h]
$$\begin{array}{c}
\includegraphics[width=8cm,keepaspectratio]{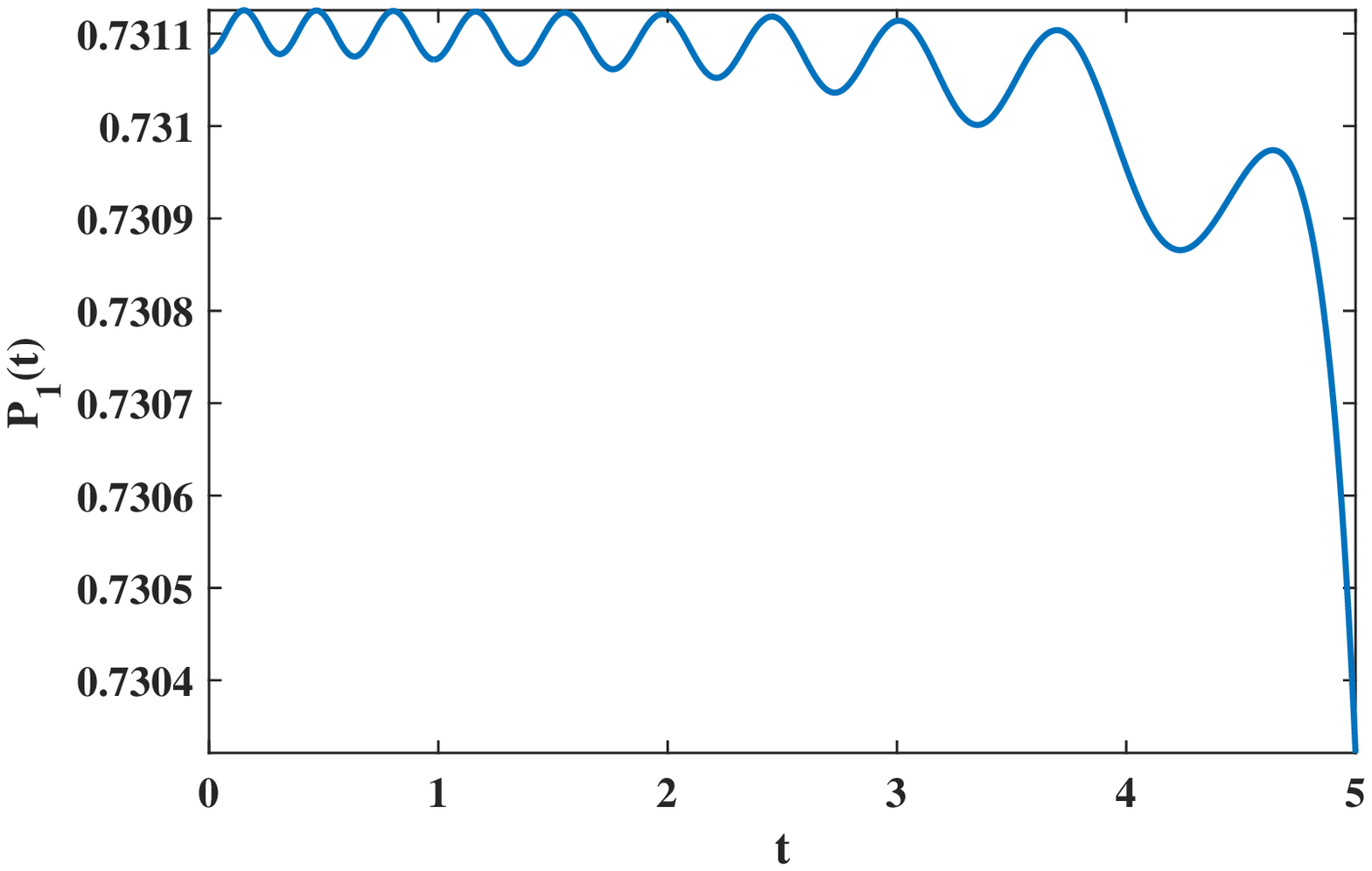}\\
\includegraphics[width=8cm,keepaspectratio]{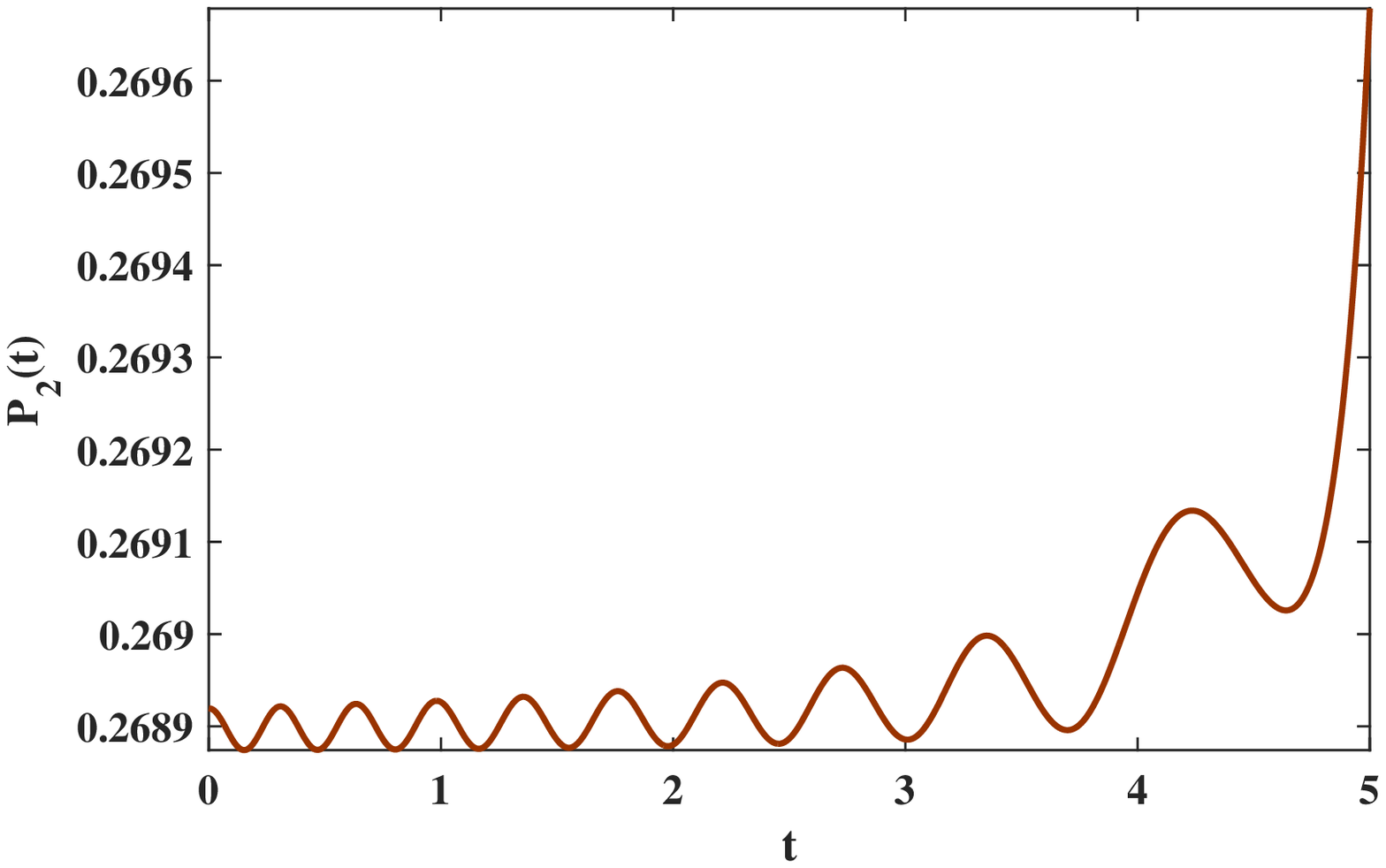}
\end{array}$$
\caption{Variation of the probabilities $P_1$ and $P_2$ of the system eigenstates during the adiabatic evolutions. We have considered here $\tau=5$ and $B_L=1$, while the other parameters are the same as in Fig. 2. Clearly these probabilities do not change substantially, as expected in an adiabatic process, though the energy eigenvalues do change.}
\end{figure}

In this regard, we further analyze the validity of the adiabaticity condition in quantum mechanics that reads as
$\left|\left\langle E_1|\frac{dE_{2}}{dt}\right\rangle \right|\ll\frac{\left|E_{1}-E_{2}\right|}{\hbar}$ \cite{messaiah}. In the present case, the adiabatic strokes are performed by changing the magnetic field strength $B$ only. Using Eqs. (\ref{en}), we obtain the following condition for adiabaticity:

\begin{equation}\label{cond1}
\xi=\left|\frac{\dot{B}g^2}{2N_+N_-(g^2+B^2)}\right|=\left|\frac{\dot{B}g}{4(g^2+B^2)^{3/2}}\right|\ll 1\;,
\end{equation}
where $\xi$ is the adiabaticity parameter and $N_\pm$ are given by Eq. (\ref{npm}). In Fig. 4, we show how the condition of adiabaticity is satisfied for different choices of the rate of change $\dot{B}$ of the magnetic field. Clearly, the adiabaticity is maintained, even for the evolution for a finite time $\tau$.
\begin{figure}
\includegraphics[width=8cm,height=6cm]{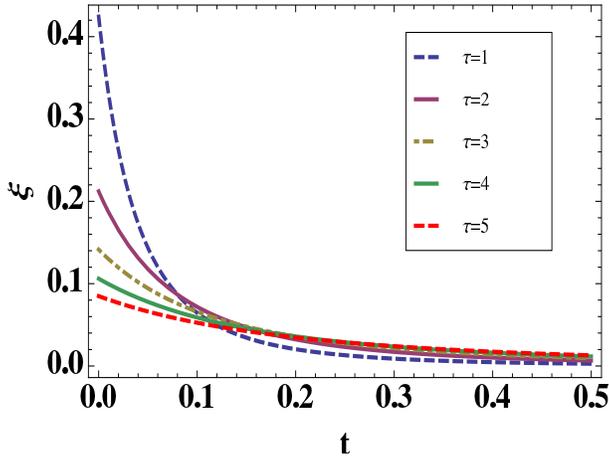}
\caption{Variation of the adiabaticity parameter $\xi$ with respect to time $t$. The parameters chosen are $B_H=10$, $B_L=1$, and $g=0.2$. Clearly $\xi$ remains much less than unity for all the times for larger $\tau$ (and therefore for slower change of the magnetic field), thereby referring to an adiabatic evolution.}
\end{figure}

In the case of $g\ll B(t)$ for all the times $t$, we can rewrite the adiabaticity parameter $\xi$ as

\begin{equation}\label{small-g}
\xi\approx \frac{\dot{B}g}{4B^3}\;.
\end{equation}
For a change of magnetic field from $B_H$ to $B_L$ in a finite time interval $\tau$, we obtain the following condition for adiabaticity:

\begin{equation}\label{cond3}
\tau \gg \left|\frac{g}{8}\left(\frac{1}{B_H^2}-\frac{1}{B_L^2}\right)\right|\;.
\end{equation}
It must be reminded that to obtain larger efficiency, one needs to have $B_L\rightarrow 0$ adiabatically [see  Eq. (\ref{eff-fin})]. Therefore, as clear from Eq. (\ref{cond3}), the corresponding time-interval becomes quite large.

\section{Conclusion}\label{s:v}
We have shown how a single trapped ion can be used to implement all the heat strokes of a quantum Otto engine. The electronic degree of freedom of the ion can be considered as the working fluid, that interacts with the thermal bath and the vibrational degree of freedom. In the Lamb-Dicke limit, the vibrational mode is confined to its two  lowest eigenstates and therefore can be considered as a cold bath. We show that by adiabatically changing the local magnetic field, the work efficiency can be made close to unity. The interaction with the hot and the cold bath is never switched off. In this framework, we show that the heat release to the cold bath can be mimicked by a projective measurement of the electronic state of the ion. We emphasize that the present proposal can be implemented using the current trapped-ion technology, routinely used for quantum computing.


\begin{thebibliography}{50}
\bibitem{scovil}\Name{Scovil H. E. D. \and Schulz-DuBois E. O.} \REVIEW{Phys. Rev.
Lett.}{2}{1959}{262}.
\bibitem{Scully-science}\Name{Scully M. O., Zubairy M. S., Agarwal G. S. 
\and Walther H.} \REVIEW{Science}{299}{2003}{862}.
\bibitem{scully-PRL2}\Name{Scully M. O.} \REVIEW{Phys. Rev. Lett.}{88}{2002}{050602}.
\bibitem{Scully-PRL-negentropy}\Name{Scully M. O.} \REVIEW{Phys. Rev. Lett.}{87}{2001}{220601}.
\bibitem{quan}\Name{Quan H. T., Liu Y. X., Sun C. P. \and Nori F.} \REVIEW{Phys. Rev. E}{76}{2007}{031105}; \Name{Quan H. T., Zhang P. \and Sun C. P.}
\REVIEW{Phys. Rev. E}{73}{2006}{036122}.
\bibitem{kosloff}\Name{Feldmann T. \and Kosloff R.} \REVIEW{Phys. Rev. E}{61}{2000}{4774}; \REVIEW{Phys. Rev. E}{68}{2003}{016101}; \REVIEW{Phys. Rev. E}{70}{2004}{046110}.

\bibitem{he}\Name{He J. Z., Chen J. C. \and Hua B.} \REVIEW{Phys. Rev. E}{65}{2002}
{036145}.

\bibitem{wu}\Name{Wu F., Chen L. G., Sun F. R., Wu C. \and Li Q.} \REVIEW{Phys.
Rev. E}{73}{2006}{016103}.


\bibitem{wang1}\Name{Wang J., Wu Z. \and He J.} \REVIEW{Phys. Rev. E}{85}{2012} {041148}.
\bibitem{rezek}\Name{Rezek Y. \and Kosloff R.} \REVIEW{New J. Phys.}{8}{2006}{83}.

\bibitem{bender}\Name{Bender C. M., Brody D. C. \and Meister B. K.} \REVIEW{J. Phys. A}{33}{2000}{4427}; \REVIEW{Proc. R. Soc. London, Ser. A}{458}{2002}{1519}.

\bibitem{henrich}\Name{Henrich M. J., Mahler G. \and Michel M.} \REVIEW{Phys. Rev. E}{75}{2007}{051118}.

\bibitem{abe}\Name{Abe S. \and  Okuyama S.} \REVIEW{Phys. Rev. E} {83} {2011} {021121}.

\bibitem{quan1}\Name{Quan H. T.} \REVIEW{Phys. Rev. E} {79} {2009} {041129}.


\bibitem{must1}\Name{Atlintas F. \and Mustecaplioglu O. E.} \REVIEW{Phys. Rev. E} {92} {2015} {022142}.

\bibitem{johal}\Name{Thomas G. and Johal R. S.}  \REVIEW{Phys. Rev. E} {83} {2011} {031135}.

\bibitem{zhang}\Name{Zhang T., Liu W.-T., Chen P.-X. \and Li C.-Z.} \REVIEW{Phys. Rev. A} {75} {2007} {062102}; \Name{ J.-Zhou He, Xian He \and  Jie Z.} \REVIEW{Chin. Phys. B} {21} {2012} {050303}.

\bibitem{albayrak} \Name{Albayrak E.} \REVIEW{Int. J. Quant. Inf.} {11} {2013}{1350021}.

\bibitem{must2} \Name{Atlintas F. , Hardal A. U. C. \and Mustecaplioglu O. E.} \REVIEW{Phys. Rev. E} {90} {2014} {032102}.

\bibitem{zhang1} \Name{Zhang G.-F.} \REVIEW{Eur. Phys. J. D} {49} {2008} {123}.

\bibitem{wang} \Name{Wang J. H., He J. Z. \and He X.} \REVIEW{Phys. Rev.
E} {84} {2011} {041127}; \Name{Wang J. H. \and  He J. Z.} \REVIEW{J. Appl. Phys.} {11} {2012} {043505}.


\bibitem{hardal} \Name{Hardal A. U. C. \and Mustecaplioglu O. E.} \REVIEW{Sci. Rep.} {5} {2015} {12952}.

\bibitem{kurizki} \Name{Klimovsky D. Gelbwaser-,Niedenzu  W. \and Kurizki G.} \REVIEW{Advances In Atomic, Molecular, and Optical Physics} {64} {2015} {329} \and references therein.

\bibitem{kurizki1}\Name{Klimovsky D. Gelbwaser-, Alicki R. \and Kurizki G.} \REVIEW{Phys. Rev. E} {87} {2013} {012140}; \Name{Klimovsky D. Gelbwaser-, Alicki R. \and Kurizki G.} \REVIEW{Europhys. Lett.} {103} {2013} {60005};  \Name{Klimovsky D. Gelbwaser- \and Kurizki G.} \REVIEW{Phys. Rev. E} {90} {2014} {022102}; \REVIEW{Sci. Rep.} {5} {2015} {7809}; \Name{Kosloff R. \and Levy A.} \REVIEW{Annu. Rev. Phys. Chem} {65} {2014} {365}.

\bibitem{bang} \Name{Viola L.  \and Lloyd S.} \REVIEW{Phys. Rev. A} {58} {1998} {2733}; \Name{Vitali D.  and Tombesi P.} \REVIEW{Phys. Rev. A} {59} {1999} {4178}.

\bibitem{abah} \Name{Abah O., Ro\ss�nagel J., Jacob  G., Diffner S., Schmidekaler F., Singer  K. \and Lutz E.} \REVIEW{Phys Rev. Lett.} {109} {2012} {203006}.

\bibitem{ion-expt} \Name{Ro\ss�nagel J., Dawkins S. T., Tolazzi K. N., Abah O., Lutz E., Schmidt-Kaler F. \and Singer K.} \REVIEW{Science} {352} {2016} {325}.

\bibitem{ion-review} \Name{Leibfried D., Blatt R., Monroe C. \and Wineland D.} \REVIEW{Rev. Mod. Phys.} {75} {2003} {281}.

\bibitem{kieu} \Name{Kieu T. D.} \REVIEW{Phys. Rev. Lett.} {93} {2004} {140403}.

\bibitem{alg} \Name{Brassard, G., Elias, Y.,  Mor, T. \and Weinstein, Y.} \REVIEW{The European Physical Journal Plus.} {129} {2014} {258}.

\bibitem{oscar} \Name{Boykin P. Oscar, Mor  T., Roychowdhury  V., Vatan F. \and Vrijen R.}, \REVIEW{PNAS} {99} {2002} {3388}; \Name{Fernandez Jose M., Lloyd S., Mor T., Roychowdhury V.} \REVIEW{International Journal of Quantum Information}, {02} {2004} {461}.

\bibitem{meas} \Name{Erez Noam, Gordon Goren, Nest Mathias \and Kurizki Gershon} \REVIEW{Nature} {452} {2008} {724}; \Name{Gordon Goren, Bensky Guy, Gelbwaser-Klimovsky David, Rao D D Bhaktavatsala, Erez Noam  \and Kurizki  Gershon} \REVIEW{New J. Phys.} {11} {2009} {123025}; \Name{Klimovsky D. Gelbwaser-, Erez N., Alicki  R. \and Kurizki G.} \REVIEW{Phys. Rev. A} {88} {2013} {022112}.

\bibitem{zeem} \Name{Zemansky M. W. \and Dittman R. H.} \Book{Heat and Thermodynamics} \Publ{McGraw Hill,7th Ed.,}  \Year{1997}.

\bibitem{goold}\Name{Goold J., Huber M., Riera A.,
del Rio L., \and Skrzypczyk P.}\REVIEW{J. Phys. A} {49} {2016}{143001}; \Name{Binder F., Vinjanampathy S., Modi K. \and Goold J.}\REVIEW{Phys. Rev. E} {91} {2015} {032119}.

\bibitem{monroe} \Name{Monroe C. {\it et. al.}} \REVIEW{Phys. Rev. Lett} {75} {1995} {4011}.

\bibitem{asoka} \Name{Biswas A. \and Brumer P.} \REVIEW{Israel J. Chem} {52} {2012} {461}.


\bibitem{messaiah} \Name{Messiah A.} \Book{Quantum Mechanics} \Publ{Dover} \Year{2014}.

\bibitem{abdel} \Name{Abdelkhalek K., Nakata  Y. \and Reeb D.} {\it arXiv:1609.06981}.
\bibitem{anders}\Name{Kammerlander P. \and Anders J.} \REVIEW{Scientific Reports}{6}{2016}{22174}.







\end{thebibliography}
\end{document}